\documentstyle[12pt]{article}
\def\NEG#1{{\rlap/#1}}

\begin{document}

\author{C.\ Bizdadea, S. O. Saliu\thanks{%
e-mail address: odile.saliu@comp-craiova.ro} and E. N. T\^\i mneanu \\
Department of Physics, University of Craiova\\
13 A.\ I.\ Cuza Str., Craiova R-1100, Romania}
\title{On the Irreducible BRST Quantization of Spin-5/2 Gauge Fields }
\maketitle

\begin{abstract}
Spin-5/2 gauge fields are quantized in an irreducible way within both the
BRST and BRST-anti-BRST manners. To this end, we transform the reducible
generating set into an irreducible one, such that the physical observables
corresponding to these two formulations coincide. The gauge-fixing procedure
emphasizes on the one hand the differences among our procedure and the
results obtained in the literature, and on the other hand the equivalence
between our BRST and BRST-anti-BRST approaches.

\end{abstract}

\section{Introduction}

The power of the antifield BRST formalism \cite{1}-\cite{5} has been fully
proved lately. This approach can be applied to both irreducible and
reducible theories. A more symmetrical treatment is given by the antifield
BRST-anti-BRST method \cite{6}-\cite{16}. Although less important than the
BRST symmetry, the BRST-anti-BRST procedure helps at a correct understanding
of the non-minimal sector. The non-minimal variables are particularly
important when dealing with redundant systems, being required during the
gauge-fixing process. A typical class of reducible models are free massless
higher spin gauge fields \cite{17}-\cite{28}. Such theories are important
due to their connection with string theory, and, because of their remarkable
gauge symmetries, they are promising candidates for building a unified
physical theory. In the meantime, the existence of a large class of
nontrivial interacting higher spin gauge theories \cite{29}, at least in
four dimensions, reveal the necessity of investigating this type of models.

In this paper we quantize free massless spin-5/2 gauge fields. Although
first-stage reducible, we show that this model can be consistently
approached in an irreducible manner following both antifield BRST and
BRST-anti-BRST lines. As far as we know, there has not been published such a
procedure. Our analysis mainly consists in: (i) replacing the reducible
generating set of the original gauge symmetries with an irreducible one, and
(ii) quantizing the irreducible theory. The irreducible model is obtained by
introducing a spin-1/2 gauge field such that the physical observables
arising from the reducible, respectively, irreducible situation are the
same. We mention that the idea of replacing the reducible symmetry by an
irreducible one acting on new variables is not new. In fact, it originates
in the Hamiltonian formalism, where a reducible set of first-class
constraints can be replaced with an irreducible one via introducing some new
variables \cite{30}-\cite{31}.

The paper is structured in five sections. In section 2, we give a brief
description of the model under study. Sections 3 and 4 are devoted to the
irreducible BRST, respectively, BRST-anti-BRST quantization. Section 5
presents some final conclusions.

\section{Spin-5/2 gauge fields}

We start with the Lagrangian action \cite{18}, \cite{24} 
\begin{eqnarray}\label{1}
& &S_0^L\left[ \psi _{\mu \nu }\right] =\int d^4x\left( -\frac 12\bar 
\psi _{\mu \nu }\NEG{\partial }\psi _{\mu \nu }-\bar \psi _{\mu \nu }
\gamma _\nu \NEG{\partial } \gamma _\lambda \psi _{\lambda \mu }+
2\bar \psi _{\mu \nu }\gamma _\nu \partial _\lambda \psi _{\lambda \mu }+
\right. \nonumber \\
& &\left. \frac 14\bar \psi _{\lambda \lambda }\NEG{\partial }\psi _{\mu \mu }-
\bar \psi _{\lambda \lambda }\partial _\mu 
\gamma _\nu \psi _{\mu \nu }\right) , 
\end{eqnarray} where $\psi _{\mu \nu }$ is a symmetric Majorana spin tensor.
In the sequel, we work with the Pauli metric ($\mu =1,2,3,4$) and Hermitian $%
\gamma $-matrices satisfying 
\begin{equation}
\label{2}\gamma _\mu \gamma _\nu +\gamma _\nu \gamma _\mu =2\delta _{\mu \nu
}. 
\end{equation}
Action (\ref{1}) is invariant under the gauge transformations 
\begin{equation}
\label{3}\delta _\epsilon \psi _{\mu \nu }=\left( \delta _{\nu \beta
}\partial _\mu +\delta _{\mu \beta }\partial _\nu \right) \left( \delta
_{\beta \alpha }-\frac 14\gamma _\beta \gamma _\alpha \right) \epsilon
_\alpha \equiv Z_{\mu \nu \alpha }\epsilon _\alpha , 
\end{equation}
with $\epsilon _\alpha $ independent gauge parameters. The transformations (%
\ref{3}) are first-stage reducible \cite{2} 
\begin{equation}
\label{4}Z_{\mu \nu \alpha }Z_\alpha =0, 
\end{equation}
with the reducibility functions 
\begin{equation}
\label{5}Z_\alpha =\gamma _\alpha . 
\end{equation}
This completes the classical Lagrangian analysis.

\section{The irreducible BRST treatment}

In this section we develop an irreducible antifield BRST approach for the
spin-5/2 gauge fields. Initially, we transform the reducible gauge
generators $Z_{\mu \nu \alpha }$ into some irreducible ones. To this end, we
associate a spin-1/2 field, $\varphi $, with the reducibility relation (\ref
{4}) and impose its gauge transformation as 
\begin{equation}
\label{6}\delta _\epsilon \varphi =A_\alpha \epsilon _\alpha , 
\end{equation}
with $A_\alpha $ some matrices (that may involve the fields) taken to fulfil 
\begin{equation}
\label{7}\det \left( A_\alpha Z_\alpha \right) \neq 0. 
\end{equation}
From (\ref{5}) and (\ref{7}) one can easily see that a possible choice reads
as 
\begin{equation}
\label{8}A_\alpha ={\bf 1}\partial _\alpha , 
\end{equation}
because $A_\alpha Z_\alpha =\NEG{\partial } $ has the inverse $\NEG{%
\partial } /\Box $. In this way, (\ref{6}) become 
\begin{equation}
\label{9}\delta _\epsilon \varphi =\partial _\alpha \epsilon _\alpha . 
\end{equation}

Next, we investigate the theory described by the action 
\begin{equation}
\label{10}S_0^L\left[ \psi _{\mu \nu },\varphi \right] =S_0^L\left[ \psi
_{\mu \nu }\right] , 
\end{equation}
subject to the gauge transformations (\ref{3}) and (\ref{9}). A noteworthy
feature of this theory is that its gauge transformations are irreducible on
account of (\ref{7}). It is remarkable that the physical observables
corresponding to the irreducible, respectively, reducible models coincide.
This can be seen as follows. Let $F\left( \psi _{\mu \nu },\varphi \right) $
be an observable of the irreducible theory. Then, its gauge variation should
vanish (at least when the equations of motion hold). This implies 
\begin{equation}
\label{11}\frac{\delta F}{\delta \psi _{\mu \nu }}Z_{\mu \nu \alpha }+\frac{%
\delta F}{\delta \varphi }A_\alpha =0. 
\end{equation}
Multiplying (\ref{11}) by $Z_\alpha $, and using (\ref{4}) and (\ref{7}) we
find 
\begin{equation}
\label{12}\frac{\delta F}{\delta \varphi }=0. 
\end{equation}
From (\ref{11}) and (\ref{12}) we obtain 
\begin{equation}
\label{13}\frac{\delta F}{\delta \psi _{\mu \nu }}Z_{\mu \nu \alpha }=0. 
\end{equation}
The last formula shows that if $F$ is an observable for the irreducible
theory, it is observable also for the reducible one. Conversely, if $\bar
F\left( \psi _{\mu \nu }\right) $ is an observable for the reducible model
(i.e. $\bar F$ verifies (\ref{13})), then it remains so in the irreducible
case because it automatically checks (\ref{11}). In conclusion, the zeroth
order cohomological groups associated with the irreducible, respectively,
reducible BRST operator are equal. Moreover, the numbers of physical degrees
of freedom corresponding to both cases are equal, therefore the path
integrals associated with the irreducible and reducible systems describe the
same theory.

It is well-known that the BRST construction relies on homological
perturbation theory that requires the acyclicity of the Koszul-Tate
operator, $\delta $. For a given gauge theory, $\delta $ can be recursively
derived antighost level by antighost level. In our irreducible approach, the
minimal ghost spectrum contains the bosonic ghosts $\eta _\alpha $ with
ghost number one, while the minimal antifield spectrum involves the fields 
\begin{equation}
\label{14}\left( \psi _{\mu \nu }^{*},\varphi ^{*},\eta _\alpha ^{*}\right)
, 
\end{equation}
with the ghost numbers ($gh$) and Grassmann parities ($\epsilon $) expressed
by 
\begin{equation}
\label{15}gh\left( \psi _{\mu \nu }^{*}\right) =gh\left( \varphi ^{*}\right)
=-1,\;gh\left( \eta _\alpha ^{*}\right) =-2, 
\end{equation}
\begin{equation}
\label{16}\epsilon \left( \psi _{\mu \nu }^{*}\right) =\epsilon \left(
\varphi ^{*}\right) =0,\;\epsilon \left( \eta _\alpha ^{*}\right) =1. 
\end{equation}
We define the action of $\delta $, as usually, through 
\begin{equation}
\label{17}\delta \psi _{\mu \nu }=0,\;\delta \varphi =0,\;\delta \eta
_\alpha =0, 
\end{equation}
\begin{equation}
\label{18}\delta \psi _{\mu \nu }^{*}=-\frac{\delta S_0^L}{\delta \psi _{\mu
\nu }},\;\delta \varphi ^{*}=-\frac{\delta S_0^L}{\delta \varphi }=0, 
\end{equation}
\begin{equation}
\label{19}\delta \eta _\alpha ^{*}=2\left( \delta _{\alpha \beta }-\frac
14\gamma _\alpha \gamma _\beta \right) \partial _\mu \psi _{\mu \beta
}^{*}+\partial _\alpha \varphi ^{*}. 
\end{equation}
The antifield $\varphi ^{*}$ being $\delta $-closed, it follows that there
can be nontrivial co-cycles in the homology of $\delta $ at non-vanishing
resolution degrees. In order to show that $\delta $ is however acyclic, we
prove that $\varphi ^{*}$ is $\delta $-exact. Multiplying (\ref{19}) from
the left by $\gamma _\alpha $, we find after simple computation 
\begin{equation}
\label{20}\varphi ^{*}=\delta \left( \frac{\NEG{\partial }}{\Box }\gamma
_\alpha \eta _\alpha ^{*}\right) , 
\end{equation}
hence $\varphi ^{*}$ is $\delta $-exact.

The last step of this treatment resides in deriving the path integral of the
irreducible theory. With the above spectra at hand, the non-minimal solution
of the master equation reads as 
\begin{eqnarray}\label{21}
& &S=S_0^L+\int d^4x\left( \bar \psi _
{\mu \nu }^{*}\left( \delta _{\nu \beta }
\partial _\mu +\delta _{\mu \beta }\partial _\nu \right) 
\left( \delta _{\beta \alpha }-\frac 14\gamma _\beta \gamma _\alpha \right) 
\eta _\alpha +\right. \nonumber \\
& &\left. \bar \varphi ^{*}\partial _\alpha \eta _\alpha +\bar b_\alpha 
{\cal{C}}_\alpha ^{*}\right) , 
\end{eqnarray} where any bar variable denotes the conjugated of the
corresponding field, and $\left( b_\alpha ,b_\alpha ^{*},{\cal C}_\alpha ,%
{\cal C}_\alpha ^{*}\right) $ form the non-minimal sector.

We pass to the gauge-fixing procedure. For subsequent purpose, we will
implement three gauge-fixing fermions. The first fermion is taken of the
form 
\begin{equation}
\label{22}K=\int d^4x\overline{{\cal C}}_\alpha \chi _\alpha ,
\end{equation}
with 
\begin{equation}
\label{23}\chi _\alpha =\gamma _\nu \psi _{\alpha \nu }-\frac 14\gamma
_\alpha \psi _{\nu \nu }+\gamma _\alpha \varphi -\frac 12b_\alpha \equiv
\rho _\alpha -\frac 12b_\alpha .
\end{equation}
Eliminating the antifields from (\ref{21}) with the aid of (\ref{22}), we
arrive at the gauge-fixed action 
\begin{eqnarray}\label{24}
& &S_K=S_0^L+\int d^4x\left( \overline{{\cal{C}}}_\alpha \gamma _\alpha 
\partial _\mu \eta _\mu +\bar b_\alpha \chi _\alpha +\right. \nonumber \\
& &\left. \frac 12\overline{{\cal{C}}}_\lambda \left( \delta _{\lambda \mu }
\gamma _\nu +\delta _{\lambda \nu }\gamma _\mu -\frac 12\delta _{\mu \nu }
\gamma _\lambda \right) \left( \delta _{\nu \beta }\partial _\mu +
\delta _{\mu \beta }\partial _\nu \right) \left( \delta _{\beta \alpha }-
\frac 14\gamma _\beta \gamma _\alpha \right) \eta _\alpha \right) . 
\end{eqnarray} Using the concrete form of (\ref{24}), we can emphasize
clearer the advantages of our irreducible procedure. The last term from (\ref
{24}) is invariant (as in the reducible case) under the gauge
transformations 
\begin{equation}
\label{25}\overline{{\cal C}}_\lambda \rightarrow \overline{{\cal C}}%
_\lambda +\overline{{\cal C}}\gamma _\lambda ,\;\eta _\alpha \rightarrow
\eta _\alpha +\gamma _\alpha \eta ,
\end{equation}
with $\overline{{\cal C}}$ and $\eta $ arbitrary spinors. These invariances
are however cancelled by the term $\overline{{\cal C}}_\alpha \gamma _\alpha
\partial _\mu \eta _\mu $, which simultaneously fixes $\overline{{\cal C}}%
_\lambda $ and $\eta _\alpha $. In the reducible approach, there is
necessary to supplement the ghost spectrum with ghosts of ghosts (and
consequently enlarge the non-minimal sector) in order to fix the above
invariances, in contrast with the present case. In order to make the link
with the reducible approach exposed in \cite{2}, we choose an alternative
gauge-fixing fermion under the form 
\begin{equation}
\label{26}K^{\prime }=a\int d^4x\overline{{\cal C}}_\alpha \NEG{\partial }%
\chi _\alpha ,
\end{equation}
with $a$ a non-vanishing constant. In this case the corresponding
gauge-fixed action is given by 
\begin{eqnarray}\label{27}
& &S_{K^{\prime }}=S_0^L+a\int d^4x\left( \overline{{\cal{C}}}_\alpha \NEG{%
\partial }\gamma _\alpha \partial _\mu \eta _\mu +\bar b_\alpha \NEG{%
\partial }\chi _\alpha +\right. \nonumber \\
& &\left. \overline{{\cal{C}}}_\lambda \NEG{\partial }\left( \delta _
{\lambda \mu }\gamma _\nu +\delta _{\lambda \nu }\gamma _\mu -
\frac 12\delta _{\mu \nu }\gamma _\lambda \right) 
\left( \delta _{\nu \alpha }-
\frac 14\gamma _\nu \gamma _\alpha \right) 
\partial _\mu \eta _\alpha \right) . 
\end{eqnarray} Eliminating the auxiliary fields $b_\alpha $ on their
equations of motion, we find 
\begin{eqnarray}\label{28}
& &S_{K^{\prime }}=S_0^L+a\int d^4x\left( \overline{{\cal{C}}}_\alpha \NEG{%
\partial }\gamma _\alpha \partial _\mu \eta _\mu +\frac 12\bar \rho _
\alpha \NEG{\partial }\rho _\alpha +\right. \nonumber \\
& &\left. \overline{{\cal{C}}}_\lambda \NEG{\partial }\left( \delta _
{\lambda \mu }\gamma _\nu +\delta _{\lambda \nu }\gamma _\mu -
\frac 12\delta _{\mu \nu }\gamma _\lambda 
\right) \left( \delta _{\nu \alpha }-
\frac 14\gamma _\nu \gamma _\alpha \right) 
\partial _\mu \eta _\alpha \right) . 
\end{eqnarray} The last two terms in (\ref{28}) are identical with the
corresponding ones from \cite{2}, while the term $\overline{{\cal C}}_\alpha 
\NEG{\partial }\gamma _\alpha \partial _\mu \eta _\mu $ replaces the
remaining terms appearing in $S_{{\rm gauge}}$ and $S_{{\rm ghost}}$ derived
within this reference. The ``Nielsen-Kallosh ghost'' for spin-5/2 gauge
fields present in \cite{2} is absent in our procedure, but the role of the
extraghost $C_1^{\prime }$ is played here by $\varphi $. The third
gauge-fixing fermion to be discussed below allows us to make a proper
correlation with the gauge-fixed action to be derived in the framework of
the BRST-anti-BRST formalism (see the next section). It has the expression 
\begin{equation}
\label{29}K^{\prime \prime }=\int d^4x\overline{{\cal C}}_\alpha \psi
_\alpha ,
\end{equation}
with 
\begin{equation}
\label{30}\psi _\alpha =2\left( \delta _{\alpha \beta }-\frac 14\gamma
_\alpha \gamma _\beta \right) \partial _\beta \varphi +\partial _\alpha \psi
_{\lambda \lambda }+\NEG{\partial }b_\alpha .
\end{equation}
The resulting gauge-fixed action is 
\begin{eqnarray}\label{31}
& &S_{K^{\prime \prime }}=S_0^L+\int d^4x\left( -2\left( \partial _\alpha 
\overline{{\cal{C}}}_\alpha \right) \left( \delta _{\beta \lambda }-
\frac 14\gamma _\beta \gamma _\lambda \right) \partial _\beta \eta _\lambda -
\right. \nonumber \\
& &\left. 2\left( \partial _\beta \overline{{\cal{C}}}_\alpha 
\right) \left( \delta _
{\beta \alpha }-\frac 14\gamma _\alpha \gamma _\beta \right) \partial _
\lambda \eta _\lambda +\bar b_\alpha \psi _\alpha \right) . 
\end{eqnarray} The gauge conditions implemented via (\ref{30}) have been not
used so far in the literature. Nevertheless, (\ref{30}) stand for some good
canonical gauge conditions because they lead to the terms 
\begin{equation}
\label{32}\frac 12\int d^4x\left( \bar \psi _{\mu \mu }\NEG{\partial }\psi
_{\nu \nu }+\frac 32{\bar \varphi \NEG{\partial }\varphi }+\bar \psi _{\mu \mu
}\NEG{\partial }\varphi \right) ,
\end{equation}
in the gauge-fixed action after eliminating the auxiliary fields $b_\alpha $
(on their field equations). All the terms in (\ref{32}) are linear in the
derivatives, as requested by field theories with fermions in order to
prevent the existence of negative-norm states.

\section{The irreducible BRST-anti-BRST procedure}

Here we develop the antifield BRST-anti-BRST quantization of action (\ref{10}%
), subject to the irreducible gauge transformations (\ref{3}) and (\ref{9}).
In connection with the general approach of the antifield BRST-anti-BRST
treatment, we follow the line from \cite{6}, \cite{10}. However, the ideas
from \cite{6}, \cite{10} are not enough in the context of our irreducible
procedure. They have to be supplemented with the analysis from the beginning
of section 3. The field, respectively, ghost spectra read as 
\begin{equation}
\label{33}\left( \stackrel{(0,0)}{\psi _{\mu \nu }},\stackrel{(0,0)}{\varphi 
}\right) , 
\end{equation}
\begin{equation}
\label{34}\left( \stackrel{(1,0)}{\eta _{1\alpha }},\stackrel{(0,1)}{\eta
_{2\alpha }},\stackrel{(1,1)}{\pi _\alpha }\right) , 
\end{equation}
while the antifield spectrum is given by 
\begin{equation}
\label{35}\left( \stackrel{(-1,0)}{\psi _{\mu \nu }^{*(1)}},\stackrel{(-1,0)%
}{\varphi _{}^{*(1)}},\stackrel{(0,-1)}{\psi _{\mu \nu }^{*(2)}},\stackrel{%
(0,-1)}{\varphi _{}^{*(2)}}\right) , 
\end{equation}
\begin{equation}
\label{36}\left( \stackrel{(-2,0)}{\eta _{1\alpha }^{*(1)}},\stackrel{(-1,-1)%
}{\eta _{2\alpha }^{*(1)}},\stackrel{(-1,-1)}{\eta _{1\alpha }^{*(2)}},%
\stackrel{(0,-2)}{\eta _{2\alpha }^{*(2)}},\stackrel{(-1,-1)}{\psi _{\mu \nu
}^{(B)}},\stackrel{(-1,-1)}{\varphi _{}^{(B)}}\right) , 
\end{equation}
\begin{equation}
\label{37}\left( \stackrel{(-2,-1)}{\pi _\alpha ^{*(1)}},\stackrel{(-1,-2)}{%
\pi _\alpha ^{*(2)}},\stackrel{(-2,-1)}{\eta _{1\alpha }^{(B)}},\stackrel{%
(-1,-2)}{\eta _{2\alpha }^{(B)}}\right) ,\left( \stackrel{(-2,-2)}{\pi
_\alpha ^{(B)}}\right) . 
\end{equation}
In (\ref{33}--\ref{37}), the superscript $\left( a,b\right) $ denote the
bighost bidegree, the notation $F^{\left( B\right) }$ signifying the bar
variable corresponding to $F$, in order to avoid confusion with the
operation of spinor conjugation. In the sequel we will omit the superscript
for simplicity. With the help of the above spectra, we derive the solution
of the master equation in the BRST-anti-BRST formalism under the form 
\begin{eqnarray}\label{38}
& &\tilde S=S_0^L+\int d^4x\left( \bar \psi _{\mu \nu }^{*(1)}\left( \delta _
{\nu \beta }\partial _\mu +\delta _{\mu \beta }
\partial _\nu \right) \left( \delta _
{\beta \alpha }-\frac 14\gamma _\beta \gamma _\alpha 
\right) \eta _{1\alpha }+
\right. \nonumber \\
& & \bar \psi _{\mu \nu }^{*(2)}\left( \delta _{\nu \beta }\partial _\mu +
\delta _{\mu \beta }\partial _\nu \right) \left( \delta _{\beta \alpha }-
\frac 14\gamma _\beta \gamma _\alpha \right) \eta _{2\alpha }+
\bar \varphi ^{*(1)}\partial _\alpha \eta _{1\alpha }+ \nonumber \\
& &\bar \varphi ^{*(2)}\partial _\alpha \eta _{2\alpha }+
\bar \psi _{\mu \nu }^{(B)}\left( \delta _{\nu \beta }\partial _\mu +
\delta _{\mu \beta }\partial _\nu \right) \left( \delta _{\beta \alpha }-
\frac 14\gamma _\beta \gamma _\alpha \right) \pi _\alpha + \nonumber \\
& &\left. \bar \varphi ^{(B)}\partial _\alpha \pi _\alpha +
\left( \bar \eta _{1\alpha }^{*(2)}-
\bar \eta _{2\alpha }^{*(1)}\right) \pi _\alpha \right) . 
\end{eqnarray} In order to fix the gauge, we introduce the variables \cite{6}%
, \cite{10} 
\begin{equation}
\label{39}\left( \stackrel{(0,1)}{\mu _{(1)\mu \nu }^{(\psi )}},\stackrel{%
(0,1)}{\mu _{(1)}^{(\varphi )}},\stackrel{(1,1)}{\mu _{(1)\alpha }^{(\eta
_1)}},\stackrel{(0,2)}{\mu _{(1)\alpha }^{(\eta _2)}},\stackrel{(1,2)}{\mu
_{(1)\alpha }^{(\pi )}}\right) , 
\end{equation}
and consider the new solution 
\begin{eqnarray}\label{40}
& &\tilde S_1=\tilde S+\int d^4x\left( \bar \psi _{\mu \nu }^{*(2)}
\mu _{(1)\mu \nu }^{(\psi )}+\bar \varphi ^{*(2)}\mu _{(1)}^{(\varphi )}+
\bar \eta _{1\alpha }^{*(2)}\mu _{(1)\alpha }^{(\eta _1)}+\right. \nonumber \\
& &\left. \bar \eta _{2\alpha }^{*(2)}\mu _{(1)\alpha }^{(\eta _2)}+
\bar \pi _\alpha ^{*(2)}\mu _{(1)\alpha }^{(\pi )}\right) . 
\end{eqnarray} With the help of the previous solution, we can fix the gauge
taking the gauge-fixing boson 
\begin{equation}
\label{41}F=\int d^4x\left( \bar \psi _{\mu \nu }\gamma _\mu \gamma _\nu
\varphi +\bar \eta _{1\alpha }\NEG{\partial } \eta _{2\alpha }\right) . 
\end{equation}
Eliminating from (\ref{40}) the bar variables (those carrying the index $%
\left( B\right) $), and the antifields with the index $\left( 1\right) $ in
the usual way \cite{6}, \cite{10}, we obtain the gauge-fixed action 
\begin{eqnarray}\label{42}
& &\tilde S_{1F}=S_0^L+\int d^4x\left( 2\bar 
\mu _{(1)}^{(\varphi )}\left( \delta _
{\beta \alpha }-\frac 14\gamma _\beta \gamma _\alpha \right) \partial _
\beta \eta _{1\alpha }+\bar \mu _{(1)\mu \mu }^{(\psi )}\partial _\alpha 
\eta _{1\alpha }+\right. \nonumber \\
& &\bar \psi _{\mu \nu }^{*(2)}\left( \delta _{\nu \beta }\partial _\mu +
\delta _{\mu \beta }\partial _\nu \right) \left( \delta _{\beta \alpha }-
\frac 14\gamma _\beta \gamma _\alpha \right) \eta _{2\alpha }-
2\bar \varphi \left( \delta _{\beta \alpha }-
\frac 14\gamma _\beta \gamma _\alpha \right) \partial _\beta \pi _\alpha +
\nonumber \\
& &\bar \varphi ^{*(2)}\partial _\alpha \eta _{2\alpha }-
\bar \psi _{\mu \mu }\partial _\alpha \pi _\alpha +
\left( \bar \eta _{1\alpha }^{*(2)}+
\partial _\lambda \left( \bar \mu _{(1)\alpha }^
{(\eta _1)}\gamma _\lambda \right) \right) \pi _\alpha +
\bar \psi _{\mu \nu }^{*(2)}\mu _{(1)\mu \nu }^{(\psi )}+ \nonumber \\
& &\left. \bar \varphi ^{*(2)}\mu _{(1)}^{(\varphi )}+
\bar \eta _{1\alpha }^{*(2)}\mu _{(1)\alpha }^{(\eta _1)}+
\bar \eta _{2\alpha }^{*(2)}\mu _{(1)\alpha }^{(\eta _2)}+
\bar \pi _\alpha ^{*(2)}\mu _{(1)\alpha }^{(\pi )}\right) . 
\end{eqnarray} We further eliminate the auxiliary antifields with the index $%
\left( 2\right) $, and the $\mu $'s from (\ref{42}) on their equations of
motion, arriving at 
\begin{eqnarray}\label{43}
& &\tilde S_{1F}=S_0^L+\int d^4x\left( -2\left( \partial _\beta \bar \eta _
{1\alpha }\right) \left( \delta _{\alpha \beta }-
\frac 14\gamma _\alpha \gamma _\beta \right) \partial _\mu \eta _{2\mu }-
\right. \nonumber \\
& &2\left( \partial _\mu \bar \eta _{1\mu }\right) 
\left( \delta _{\beta \alpha }-
\frac 14\gamma _\beta \gamma _\alpha \right) 
\partial _\beta \eta _{2\alpha }+
\nonumber \\
& &\left. \bar \pi _\alpha \left( 2\left( \delta _{\alpha \beta }-
\frac 14\gamma _\alpha \gamma _\beta \right) \partial _\beta \varphi +
\partial _\alpha \psi _{\mu \mu }+\NEG{\partial }\pi _\alpha \right) \right) . 
\end{eqnarray} This is the final result of our irreducible antifield
BRST-anti-BRST formalism. The gauge-fixed action (\ref{43}) is identical
with (\ref{31}), modulo the identifications 
\begin{equation}
\label{44}\overline{{\cal C}}_\alpha =\bar \eta _{1\alpha },\;\eta _\alpha
=\eta _{2\alpha },\;b_\alpha =\pi _\alpha . 
\end{equation}

\section{Conclusion}

We showed that free massless spin-5/2 gauge fields can be consistently
quantized as an irreducible theory within both the antifield BRST and
BRST-anti-BRST approaches. In this context, although the starting model is
reducible, the ghosts of ghosts are not necessary. This is because we
replace the initial reducible generating set with an irreducible one, such
that the physical observables remain the same in both formulations. The
irreducibility is gained by introducing a spin-1/2 gauge field having
trivial field equation. The triviality of the spin-1/2 gauge field equation
implies the $\delta $-closedness of the associated antifield. In spite of
this, the irreducible Koszul-Tate operator is proved to be truly acyclic at
non-vanishing antighost numbers. In the framework of the BRST procedure we
discuss three possibilities of fixing the gauge. The first one emphasizes in
a clearer fashion the meaning of our irreducible treatment. The second
choice is helpful at establishing a comparison with the reducible methods
employed in the literature with regard to the investigated model. The third
election is taken in order to make manifest the equivalence with the
BRST-anti-BRST gauge-fixed action.

\end{document}